\documentclass[12pt]{article}%
\usepackage{amsmath}
\usepackage{graphicx}%
\usepackage{amsfonts}%
\usepackage{amssymb}

\begin{document}

\author{D. Galetti$^{a}$, B.M. Pimentel$^{a}$, C.L. Lima$^{b,}\thanks{Corresponding
author. E-mail address:\ cllima@if.usp.br (C.L. Lima)}$, and E.C. Silva$^a$\\$^{a}$Instituto de F\'{\i}sica Te\'{o}rica\\Universidade Estadual Paulista --- UNESP\\Rua Pamplona 145\\01405 - 900 S\~{a}o Paulo, S.P. \\Brazil\\$^{b}$Instituto de F\'{\i}sica, Universidade de S\~{a}o Paulo\\C.P. 66318, 05315-970, S\~{a}o Paulo, S.P.\\Brazil}
\title{Quantum spin tunneling of magnetization in small ferromagnetic particles}
\maketitle

\begin{abstract}
A new quantum approach is presented that can account for the description of
small ferromagnetic particles magnetization tunneling. An estimate of the
saturation value of an external applied magnetic field along the easy axis is
obtained. An analytic expression for the tunneling factor in the absence of an
external magnetic field is deduced from the present approach that also allows
to obtain the crossover temperature characterizing the regime where tunneling
is dominated by the quantum effects.

\end{abstract}

\textit{Pacs}: 75.75.+a, 73.40.Gk, 03.65.-w


\newpage

Single-domain magnetic particles constitute a scenario in which macroscopic
quantum tunneling can be studied. In such systems of mesoscopic size the
eletronic spins can form an aligned magnetic state that can be oriented in
several directions. Quantum mechanics can then be called upon to estimate the
tunneling probability that may occur, associated with the possibility of a
change of the magnetization from one direction to another. The existence of an
energy barrier to be tunneled that do not completely disappear even with a
decrease in the temperature towards absolute zero has therefore been suggested
and several attempts have been presented in order to estimate the rate of
tunneling in these cases \cite{van-suto,enz-schil,chud-gunther,scharf}. In
some approaches an adapted version of the WKB is proposed, while, from another
perspective, other approaches make use of Feynman's path integral with $su(2)$
coherent states from which the instanton method follows to treat this same
quantum problem for a single spin in predicting the energy levels splitting .

In general, those approaches start with a Hamiltonian that is associated with
a model of the physical system and some considerations are presented in order
to get a pair of complementary operators which should characterize the degree
of freedom. To this end, the spin system which is characterized by a
finite-dimensional state space, is converted into a continuous one by means of
a transformation which leads to a pair of canonical operators satisfying a new
commutation relation which is now dependent on the particle spin quantum
number \cite{villain}. It is to be observed that this construction is valid
for large spin values and that corrections of quantum character must be
introduced for small/medium values.

In the present paper we intend to show that a new quantum description recently
proposed \cite{ga5} can be used in such a way that the energy levels splitting
in ferromagnetic particles can be obtained in a direct and simple way.

We start from a model Hamiltonian describing the ferromagnetic particle
\cite{chud-gunther}
\begin{equation}
\mathcal{H}=-K_{1}S_{z}^{2}+K_{2}S_{y}^{2}+K_{3}H_{z}S_{z} \label{eq1}%
\end{equation}
written in terms of spin operators $S_{y}$ and $S_{z}$ with $K_{1},K_{2}>0$,
corresponding to the case of easy-axis anisotropy along the $z$ axis and an
external magnetic field pointing along the $z$ direction. Putting this
Hamiltonian in a form that emphasizes the creation and anihilation operators
we get
\begin{equation}
\mathcal{H}=AS_{z}^{2}+B\left(  S_{+}^{2}+S_{-}^{2}\right)  +CS^{2}+K_{3}%
H_{z}S_{z}, \label{eq1a}%
\end{equation}
where $A=-\left(  K_{1}+K_{2}/2\right)  ,\;B=-K_{2}/4,$ and $C=K_{2}/2$. From
a formal point of view, it is interesting to see that this Hamiltonian is akin
to that one suggested to describe the $Fe_{8}$ magnetic cluster (also with an
external magnetic field) and that has been treated elsewhere \cite{ga5}.

The starting point of the proposed approach is the introduction of a quantum
phenomenological Hamiltonian describing the spin system, written in terms of
angular momentum operators obeying the standard commutation relations, and
that reflects the internal symmetries of the system. It may also contain terms
taking into account external applied magnetic fields. The degree of freedom
that undergoes tunneling is considered a particular collective manifestation
of the system, and it is assumed to be the only relevant one. At the same
time, the temperature of the system is assumed so conveniently low that
possible related termally assisted processes are not taken into account so
that only quantum effects are considered.

The next step consists in getting a new Hamiltonian that is an approximate
version of Eq. (\ref{eq1a}). This new Hamiltonian is obtained through a series
of transformations performed on the matrix generated by calculating the
expectation values of Eq. (\ref{eq1a}) with the $su(2)$ coherent states
$|j,z\rangle$, where $z$ is a complex variable and $j$ characterizes the
angular momentum state multiplet \cite{perelomov}. The Hamiltonian in the
overcomplete spin coherent states representation is then given by
\begin{equation}
\langle j,z^{\prime}|H|j,z\rangle=K\left(  z^{\prime},z\right)  , \label{eq1c}%
\end{equation}
the also known generator coordinate energy kernel \cite{griffin}; it embodies
all the quantum information related to the system we want to study. The
procedure of extracting a new Hamiltonian -- written now in terms of an angle
variable -- from Eq. (\ref{eq1c}) has been already shown elsewhere and will
not be repeated here \cite{ga5,garu,gapi}. In fact, as it was proved there,
the variational generator coordinate method can in this case be used to
rewrite the Hamiltonian from which we start in an exact and discrete
representation which can then be conveniently treated in order to give an
approximate Hamiltonian in the angle representation. It is important to point
that the Hamiltonian we obtain is in fact an approximate one, but we have also
shown, by studying the Lipkin model \cite{lipkin}, that it is already a
reliable Hamiltonian for spin systems with $j\gtrsim5$ \cite{ga5}.
Furthermore, since this approach is based on quantum grounds from the
beginning, it does not need to go through any quantization process. Also, it
is not necessary to convert the discrete spin system into a continuous one as
it is usually done \cite{villain}, at the same time that the quantum character
of the angle-angular momentum pair is properly taken into account.

Now, carrying out all the steps of the approach, we obtain the approximate
Hamiltonian in the angle representation, namely
\begin{equation}
H\left(  \phi\right)  =-\frac{1}{2}\frac{d}{d\phi}\frac{1}{M\left(
\phi\right)  }\frac{d}{d\phi}+V\left(  \phi\right)  \label{eq2}%
\end{equation}
where
\begin{equation}
M\left(  \phi\right)  =\frac{1}{2\left(  K_{2}+K_{1}\cos^{2}\phi\right)
-\frac{K_{3}}{j}H_{z}\cos\phi} \label{eq3}%
\end{equation}
is the effective mass,
\begin{equation}
V\left(  \phi\right)  =-K_{1}j\left(  j+1\right)  \cos^{2}\phi+K_{3}H_{z}%
\sqrt{j\left(  j+1\right)  }\cos\phi\label{eq4}%
\end{equation}
is the potential respectively, and $j$ is the total spin of the ferromagnetic
particle. By a direct inspection we see that the potential function has minima
at $\phi=0,\pi$, while the maxima occur at
\[
\cos\phi=\frac{K_{3}H_{z}}{2K_{1}\sqrt{j\left(  j+1\right)  }},
\]
respectively. It is immediate to see that for $H_{z}=0$, the maxima are at
$\phi=\pi/2$ and $3\pi/2$.

As has been mentioned before, the accuracy of this approximation can be
checked by comparing the eigenvalues of the Schr\"{o}dinger equation
\begin{equation}
H\left(  \phi\right)  \Psi_{k}\left(  \phi\right)  =\mathcal{E}_{K}\Psi
_{k}\left(  \phi\right)  \label{eq5}%
\end{equation}
with those obtained by diagonalizing Hamiltonian (\ref{eq1}) in the $\left\{
|jm>\right\}  $ state space. By its turn, it is immediate to see that the
numerical results of Eq. (\ref{eq5}) can be directly obtained from a Fourier
analysis that give the energy eigenstates and the corresponding
eigenfunctions. It has also been shown elsewhere \cite{ga5} that already for
$j\gtrsim5$ solutions of Hamiltonians of this kind are in good agreement with
the ones coming from (\ref{eq1}), being the deviation of the order of $1\%$.

An interesting result can be immediately drawn from a direct analysis of the
potential energy expression, Eq. (\ref{eq4}). It is simple to see that the
extrema of the potential energy depend on the strength of the external
magnetic field. Furthermore, if we realise that an increase of that field
tends to gather together the two maxima, at the same time that it rises the
value of the minimum at $\phi=\pi$ (or at $\phi=0$ if the field is reversed),
we can expect to find the intensity of the field at which those three extrema
colapse into a single maximum. At this particular value of the field there is
a single minimum at $\phi=0$ (or at $\phi=\pi$ if the field is reversed) and
tunneling cannot occur. Therefore, using the proposed approach we can estimate
the value of the external magnetic field at which tunneling no longer has any
meaning if we consider the expression for the potential energy expression, Eq.
(\ref{eq4}), and look for the value of $H_{z}$ for which the minimum turns
into a maximum. After a direct calculation we obtain
\[
H_{z}^{\lim}=\frac{2K_{1}}{K_{3}}\sqrt{j\left(  j+1\right)  }%
\]
so that
\begin{equation}
H_{z}^{0}=\frac{H_{z}^{\lim}}{2j}=\frac{K_{1}}{K_{3}}\sqrt{1+\frac{1}{j}}
\label{eq13}%
\end{equation}
is the value of the external paralel magnetic field whose multiples produce a
matching of the energy levels in the two potential wells. The result obtained
from a numerical diagonalization of Eq. (\ref{eq1}) for the matching value in
a test case is in good agreement with the predicted value of Eq. (\ref{eq13}).
It is then evident that for an applied external field of magnitude
$H_{z}^{\lim}$ the resulting potential function presents only a single minimum.

It is clear that if we want to discuss all the features of the energy levels
associated with Hamiltonian (\ref{eq2}) we may numerically solve the
Schr\"{o}dinger equation (\ref{eq5}) and get the results. Although this goal
is simple and direct, it will not be pursued here.

In what concerns the energy levels splitting associated with the tunneling
through the potential barrier present in $V\left(  \phi\right)  $ in the
absence of the external magnetic field, i.e., $H_{z}=0$, we may take advantage
of the even symmetry of $V\left(  \phi\right)  $ as well as of $M\left(
\phi\right)  $ to use the well-known WKB expression ($\hbar=1)$ \cite{landau}
\begin{equation}
\Delta\mathcal{E=}\frac{\omega_{b}}{\pi}\exp\left[  -\sqrt{2\mathcal{M}}%
\int_{\phi_{i}}^{\phi_{s}}\sqrt{V\left(  \phi\right)  -\mathcal{E}}%
\;d\phi\right]  , \label{split}%
\end{equation}
where $\omega_{b}$ is the frequency associated with one of the minima of
$V\left(  \phi\right)  $, and $\mathcal{M}$ is a constant suitably taken as an
average of $M\left(  \phi\right)  $ in the barrier region. Considering
\[
M\left(  \phi_{\min}\right)  \omega_{b}^{2}=\left.  \frac{d^{2}V\left(
\phi\right)  }{d\phi^{2}}\right|  _{\phi_{\min}}%
\]
and taking into account that $\phi_{\min}=0$, $\pi$ and $\phi_{\max}=\pi/2$,
$3\pi/2$, we see that
\begin{equation}
\omega_{b}=2\sqrt{K1\left(  K1+K_{2}\right)  j\left(  j+1\right)  }.
\label{eq5b}%
\end{equation}
On the other hand, the exponential factor can be treated in an analytic way.
Taking
\[
\sqrt{-K_{1}j\left(  j+1\right)  \cos^{2}\phi+\left|  \mathcal{E}\right|
}=\sqrt{\left|  \mathcal{E}\right|  }\sqrt{1-\frac{K_{1}j\left(  j+1\right)
}{\left|  \mathcal{E}\right|  }\cos^{2}\phi}%
\]
and calling $k^{2}=K_{1}j\left(  j+1\right)  /\left|  \mathcal{E}\right|  $ we
see that the turning points of the integral are given by the equation
$\cos\phi_{ext}=\pm k^{-1}$. Therefore,
\[
I\left(  \mathcal{E}\right)  =\int_{\phi_{i}}^{\phi_{s}}\sqrt{1-k^{2}\cos
^{2}\phi}=\left[  \frac{k^{2}-1}{k}\left.  F\left(  \arcsin\left(
k\cos\right)  ,\frac{1}{k}\right)  \right|  _{\phi_{i}}^{\phi_{s}}\right.  -
\]
\[
\left.  kE\left.  \left(  \arcsin\left(  k\cos\right)  ,\frac{1}{k}\right)
\right|  _{\phi_{i}}^{\phi_{s}}\right]  \;\;\text{if }k>1,
\]
where $F\left(  \varphi,p\right)  $ and $E\left(  \varphi,p\right)  $ are the
elliptic functions of the first and second kind respectively \cite{grad}.
Since these functions are odd under the change $\varphi\rightarrow-\varphi$,
and using the expression for $\cos\phi_{i,s}$, we obtain
\begin{equation}
I\left(  \mathcal{E}\right)  =2\left[  -\frac{k^{2}-1}{k}\mathbf{K}\left(
\frac{1}{k}\right)  +k\mathbf{E}\left(  \frac{1}{k}\right)  \right]  ,
\label{eq6}%
\end{equation}
where $\mathbf{K}\left(  \frac{1}{k}\right)  =F\left(  \frac{\pi}{2}%
,\frac{1}{k}\right)  $ and $\mathbf{E}\left(  \frac{1}{k}\right)  =E\left(
\frac{\pi}{2},\frac{1}{k}\right)  $ are the complete elliptic integrals
\cite{grad}. Now, using their corresponding series representations we end up
with the result
\begin{equation}
I\left(  \mathcal{E}\right)  =\frac{\pi}{2}\left(  \frac{1}{k}+\frac{1}%
{8k^{3}}+\frac{3}{64k^{5}}+...\right)  \label{eq7}%
\end{equation}
so that, the integral to the lowest order reads
\begin{equation}
I\left(  \mathcal{E}\right)  \simeq\frac{\pi}{2k}=\frac{\pi}{2}\sqrt
{\frac{\left|  \mathcal{E}\right|  }{K_{1}j\left(  j+1\right)  }}. \label{eq8}%
\end{equation}
With that result we can then write the barrier penetrability factor
\begin{equation}
P\simeq\exp\left[  -\frac{\pi}{2}\left|  \mathcal{E}\right|  \sqrt
{\frac{2\mathcal{M}}{K_{1}j\left(  j+1\right)  }}\right]  . \label{eq9}%
\end{equation}

In order to have an estimate of the tunneling factor we have to introduce the
approximate constant mass. If we consider for instance the crude simple
arithmetic mean in order to get an analytic expression
\begin{equation}
\mathcal{M=}\frac{M\left(  \phi=0\right)  +M\left(  \phi=\frac{\pi}{2}\right)
}{2}=\frac{K_{1}+2K_{2}}{4K_{2}\left(  K_{1}+K_{2}\right)  }, \label{eq10}%
\end{equation}
for the constant mass, we obtain
\[
P\simeq\exp\left[  -\frac{\pi}{2}\left|  \mathcal{E}\right|  \sqrt
{\frac{\left(  K_{1}+2K_{2}\right)  }{2K_{1}j\left(  j+1\right)  K_{2}\left(
K_{1}+K_{2}\right)  }}\right]  .
\]
On the other hand, the frequency expression can be directly obtained from Eq.
(\ref{eq5b}) so that the expression for the energy splitting, up to first
order in the exponential, is
\begin{equation}
\Delta\mathcal{E=}\frac{2}{\pi}\sqrt{K1\left(  K1+K_{2}\right)  j\left(
j+1\right)  }\exp\left[  -\frac{\pi}{2}\left|  \mathcal{E}\right|
\sqrt{\frac{\left(  K_{1}+2K_{2}\right)  }{2K_{1}j\left(  j+1\right)
K_{2}\left(  K_{1}+K_{2}\right)  }}\right]  \label{eq11}%
\end{equation}

As it can be immediately seen, $P\left(  \mathcal{E}\right)  =1$ at the top of
the energy barrier (i.e., $\mathcal{E}=V_{\max}=0.0)$, thus we verify that
this expression is not adequate to describe tunneling occurring in energy
levels near or at the barrier top. In these cases, we may consider the
Kemble-Hill-Wheeler/Miller-Good expression \cite{k,razavy,hw,mg} for the
barrier penetration factor, namely%

\[
P\left(  \mathcal{E}\right)  =\frac{1}{1+\exp\left[  \frac{\omega_{t}}{\pi
}\left(  V_{0}-\mathcal{E}\right)  \right]  },
\]
where $\omega_{t}$ is the frequency associated with the concavity at the
potential barrier top, and $V_{0}=V\left(  \phi=\phi_{top}\right)  $. It is
important to see that at $\mathcal{E}=V_{0}$ we get $P\left(  V_{0}\right)
=0.5$ instead of the usual WKB value $P=1.0$, thus assuring that the
Kemble-Hill-Wheeler/Miller-Good expression is a better approximation for this
particular kind of situation.

If we realise that the exponential factor is the dominant part of the energy
splitting expression, Eq. (\ref{split}), we are lead to consider again
expression (\ref{eq9}) in order to estimate the temperature below which the
quantum effects dominate over the thermal ones. Since the thermal ativation is
proportional to $\exp(-V_{0}/k_{B}T)$, where $V_{0\text{ }}$is the barrier
height, and in our case it is given by $V_{0}=K_{1}j\left(  j+1\right)  $, and
considering that $\left|  \mathcal{E}\right|  \simeq K_{1}j\left(  j+1\right)
$ for $j\gg1$, then the crossover temperature between the two regimes can be
estimated by equating
\[
\frac{\pi}{2}\sqrt{\frac{2\mathcal{M}}{K_{1}j\left(  j+1\right)  }}%
=\frac{1}{k_{B}T_{c}}\text{,}%
\]
that gives
\[
T_{c}=\frac{2}{\pi k_{B}}\sqrt{\frac{K_{1}j\left(  j+1\right)  }{2\mathcal{M}%
}}\text{.}%
\]
The effective mass plays an important role in the above expression, telling us
how the transverse anisotropy governs the behaviour of the relevant physical
quantities related to the tunneling. In order to have an estimate of this
behaviour let us consider Eq. (\ref{eq10}) for the effective mass. It is
direct to see that for $K_{2}\rightarrow0$, $\mathcal{M}\rightarrow\infty$ so
that $T_{c}\rightarrow0$, and consequently $\Delta\mathcal{E}\rightarrow0$, as
expected, since then there is no tunneling at all. It is to be stressed that
the above quantitative results do not follow due to a particular form of the
effective mass used. They would also follow for any reasonable expression for
the mass. It is important to observe that the effective mass is the only term
carrying the information related to the transverse anisotropy which is
responsible for the quantum tunneling effects.\medskip

\noindent\textbf{Acknowledgements}\newline The authors would like to thank
Prof. Renato Jardim from Instituto de F\'{\i}sica da Universidade de S\~{a}o
Paulo for enlightening discussions. D.G. and B.M.P. were partially supported
by Conselho Nacional de Desenvolvimento Cient\'{\i}fico e Tecnol\'{o}gico,
CNPq, Brazil; E.C.S. was supported by CAPES, Brazil. B.M.P. and C.L.L. also
acknowledge partial support by Funda\c{c}\~{a}o de Amparo \`{a} Pesquisa do
Estado de S\~{a}o Paulo, FAPESP, Brazil.


\begin{thebibliography}{9}                                                                                                %

\bibitem {van-suto}J. L. van Hemmen and A. S\"{u}t\H{o}, Europhys. Lett
\textbf{1} (1986) 481; J. L. van Hemmen and A. S\"{u}t\H{o}, Physica
\textbf{141B} (1986) 37.

\bibitem {enz-schil}M. Enz and R. Schilling, J. Phys. \textbf{C19} (1986)
1765; M. Enz and R. Schilling, J. Phys. \textbf{C19} (1986) L711.

\bibitem {chud-gunther}E. M. Chudnovsky and L. Gunther, Phys. Rev. Lett.
\textbf{60} (1988) 661.

\bibitem {scharf}G. Scharf, W. F. Wreszinski, and J. L. van Hemmen, J. Phys.
\textbf{A20} (1987) 4309.

\bibitem {villain}J. Villain, J. de Phys. \textbf{35} (1974) 27.

\bibitem {ga5}D. Galetti, Physica \textbf{A 374} (2007) 211; doi:10.1016/j.physa.2006.07.030.

\bibitem {perelomov}A. Peremolov, \textit{Generalized Coherent States and
Their Applications, }Springer Verlag, Berlin, 1986

\bibitem {griffin}J. J. Griffin and J. A. Wheeler, Phys. Rev. \textbf{108}
(1957) 311.

\bibitem {garu}D. Galetti and M. Ruzzi, J. Phys. A: Math. Gen. \textbf{33}
(2000) 2799.

\bibitem {gapi}D. Galetti, B. M. Pimentel, and C. L. Lima, Physica
\textbf{A351} (2005) 315.

\bibitem {lipkin}H. Lipkin, N. Meshkov, and A. J. Glick, Nucl. Phys.
\textbf{62} (1965) 188.

\bibitem {landau}L. Landau, \textit{M\'{e}chanique Quantique}, MIR, 1989, Chap VII.

\bibitem {grad}I. S. Gradshteyn and I. M. Rhyzik, \textit{Table of Integrals,
Series, and Products}, Academic Press, New York, 1980.

\bibitem {k}E. C. Kemble, \textit{Fundamental Principles of Quantum
Mechanics}, McGrawHill, New York, 1937.

\bibitem {razavy}M. Razavy, \textit{Quantum Theory of Tunneling}, World
Scientific, Singapore, 2003.

\bibitem {hw}D. Hill and J. A. Wheeler, Phys. Rev. \textbf{89} (1953) 1102.

\bibitem {mg}S. C. Miller and R. M. Good, Phys. Rev. \textbf{91} (1953) 174.
\end{thebibliography}
\end{document}